\documentclass[12pt]{iopart}

\usepackage{graphicx}
\usepackage{bm}
\usepackage{iopams}
\begin{document}


\title[Electron acceleration from relativistically transparent plasma]{The Unexpected Role of Evolving Longitudinal Electric Fields in Generating Energetic Electrons in Relativistically Transparent Plasmas}

\author{L.~Willingale$^{1\ddag}$, A.~V.~Arefiev$^{2}$, G.~J.~Williams$^{3,4}$, H.~Chen$^{3}$, F.~Dollar$^{1,\dag}$, A.~U.~Hazi$^{3}$, A.~Maksimchuk$^{1}$, M.~J.-E.~Manuel$^{1,\ast}$, E.~Marley$^{3,4}$, W.~Nazarov$^{5}$, T.~Z.~Zhao$^{1}$, C.~Zulick$^{1,\ast \ast}$}
\address{$^1$ University of Michigan, 2200 Bonisteel Boulevard, Ann Arbor, Michigan 48109, USA}
\address{$^2$ Department of Mechanical and Aerospace Engineering, University of California at San Diego, La Jolla, CA 92093, USA}
\address{$^3$ Lawrence Livermore National Laboratory, Livermore, California 94550, USA}
\address{$^4$ Department of Applied Science, University of California Davis, Davis, California 95616, USA}
\address{$^5$ University of St Andrews, High Energy Laser Materials R\&D Laboratory, Unit 4, St Andrews NTC, North Haugh, St Andrews, Fife KY16 9ST, United Kingdom}

\ead{wlouise@umich.edu}
\date{\today}

\begin{abstract}
Superponderomotive-energy electrons are observed experimentally from the interaction of an intense laser pulse with a relativistically transparent target. For a relativistically transparent target, kinetic modeling shows that the generation of energetic electrons is dominated by energy transfer within the main, classically overdense, plasma volume. The laser pulse produces a narrowing, funnel-like channel inside the plasma volume that generates a field structure responsible for the electron heating. The field structure combines a slowly evolving azimuthal magnetic field, generated by a strong laser-driven longitudinal electron current, and, unexpectedly, a strong propagating longitudinal electric field, generated by reflections off the walls of the funnel-like channel. The magnetic field assists electron heating by the transverse electric field of the laser pulse through deflections, whereas the longitudinal electric field directly accelerates the electrons in the forward direction. The longitudinal electric field produced by reflections is 30 times stronger than that in the incoming laser beam and the resulting direct laser acceleration contributes roughly one third of the energy transferred by the transverse electric field of the laser pulse to electrons of the super-ponderomotive tail.\end{abstract}
\maketitle

\section{Introduction}

Electrons move and can gain energy in response to the electromagnetic fields of a laser pulse; the coupling of the laser pulse energy to the electrons regulates the entire relativistic intensity laser-plasma interaction. Many other secondary phenomena of interest arise from this electron heating, including ion acceleration \cite{ion_accn_2000,Willingale_PRL_2009,Silva_PRL_2004,Haberberger_NP_2012, Palmer_PRL_2012, Fiuza_PRL_2012, Albright_PoP_2007, Bin_PRL_2018}, high-harmonic generation \cite{HHG}, x-ray beam generation \cite{Rousse_PRL_2004,Kneip_PRL_2008,Stark_PRL_2016}, and positron production \cite{Chen_PRL_2009,Jansen_PPCF_2018}. Electron acceleration and heating in a plasma is surprisingly complex due to the collective plasma effects that affect both the laser pulse propagation and the electron motion itself.

Several parameters determine the dominant electron heating mechanism at relativistic intensities, the foremost factors being the plasma density ($n_{e}$), and the laser pulse duration and intensity. The classical critical plasma density is defined to be $n_{c} = m_{e} \epsilon_{0} \omega_{L}^{2} / e^{2}$, where $\omega_{L}$ is the laser frequency. The two extremes for target plasma densities have been studied extensively. For a very overdense ($n_{e} \gg n_{c}$), short scale-length plasma, the dominant heating mechanisms become vacuum heating \cite{Brunel_PRL_1987} and $\mathbf{j} \times \mathbf{B}$ heating \cite{Kruer_PoF_1985}, with the expected hot electron temperature scaling as the ponderomotive potential, $U_{p} \approx (a_{0} /2)^2  m_{e} c^{2}$ \cite{Wilks_PRL_1992}, where $a_{0}$ is the normalized laser amplitude. A significant scale-length underdense plasma ($n_{e} < n_{c}$) could be present ahead of an overdense target due to heating and expansion either during a laser pre-pulse or on the timescale of the laser pulse interaction~\cite{Kemp_PRL_2012, Sorokovikova_PRL_2016}. Such a pre-plasma is known to reduce the $\mathbf{j} \times \mathbf{B}$ heating and the overall energy conversion efficiency \cite{Wilks_IEEE_1997, Cai_PoP_2010,Yabuuchi_PoP_2010}. However, a characteristic enhancement in the high energy tail of escaping electrons is a typical observation from experiments~\cite{Jarrott_PoP_2014} and has been attributed to other acceleration mechanisms occurring in the underdense region~\cite{Cai_PoP_2010, Krygier_PoP_2014,Sorokovikova_PRL_2016}.
There is significant interest in using near-critical density plasma to enhance ion acceleration mechanisms \cite{Willingale_PRL_2009,Willingale_PoP_2011,Haberberger_NP_2012,Fiuza_PRL_2012,Bin_PRL_2015,Chen_SR_2017,Fedeli_SR_2018, Bin_PRL_2018}, or to generate bright x-ray \cite{Wang_PoP_2015} or electron-positron plasmas \cite{Zhu_NC_2016} by taking advantage of the high laser energy conversion to hot electrons and the high electron temperatures.

A significantly underdense plasma offers favorable conditions for electron acceleration well beyond the ponderomotive potential, as it allows the laser pulse to propagate with a phase velocity ($v_{ph}$) that remains close to the speed of light. The laser pulse could excite a co-propagating plasma wave in the underdense plasma leading to laser wake-field acceleration~\cite{Tajima_PRL_1979}. For a higher intensity laser pulse with a duration longer than a plasma wave period, the plasma wave development is inhibited due to the large and sustained ponderomotive force.
Instead, electrons are expelled from regions of highest intensity and, if the ponderomotive force persists to balance the electric field acting to return the electrons, a cavitated channel can form~\cite{Mora_PRB_1996, Tzeng_PRL_1998, Nilson_NJP_2010}. In this regime, direct laser acceleration (DLA) assisted by quasi-static transverse and longitudinal electric fields of the channel may become the dominant mechanism generating an electron population with characteristic energies many times greater than $U_{p}$~ \cite{Pukhov_PoP_1998, Pukhov_PoP_1999, MeyerterVehn_PoP_1999, Arefiev_PRL_2012, Robinson_PRL_2013, Arefiev_PoP_2014, Arefiev_PoP_2016, Khudik_PoP_2016, Jiang_PRL_2016}. 

In this paper, we consider the energy transfer mechanisms in the intermediate range of near-critical densities ($n_{e} \sim n_{c}$), a regime that has received little attention.
One compelling reason to consider near-critical density targets is that they can become transparent at relativistic laser intensities, when $a_{0} >1$. Accelerating electrons to relativistic energies, the laser pulse effectively enhances the electron mass, thus reducing the effective critical density that determines the cutoff for an electromagnetic wave. As a result, the relativistically induced transparency allows the laser pulse to propagate in plasmas with electron densities up to $n_{\gamma c} \equiv \bar{\gamma} n_{c}$~\cite{Guerin_PoP_1996, Fuchs_PRL_1998, Willingale_PRL_2009}, where $\bar{\gamma}$ is the characteristic Lorentz factor.
The expected drawback of this regime is the enhancement of $v_{ph}$ of the pulse. This superluminosity leads to poor phase matching between the wave and the electron during DLA, severely limiting the electron energy gain~\cite{Robinson_PoP_2015}.
However, the presented experimental measurements from relativistically near-critical plasma does observe an enhanced super-ponderomotive electron tail formation.
It has previously been noted that even relatively weak oscillating longitudinal electric fields found in a focussing or defocussing laser pulse can play a significant role in understanding DLA \cite{Robinson_PoP_2018}.
Here, the two-dimensional particle-in-cell simulations show one of the dominant energy transfer mechanisms into the high-energy tail is mediated by the evolving longitudinal electric fields within the main plasma volume causing the electrons to experience huge, rapid acceleration via this mechanism.
This is in stark contrast to previously identified DLA mechanisms that have either occurred in the very underdense region or essentially in vacuum with the overdense region serving as a source of electrons.

\section{Experimental set-up}
\label{exp_section}
The experiments were performed using the Titan laser system at the Jupiter Laser Facility \cite{Stuart_OSA_2006}. A pulse energy of $\epsilon_{L} = 127 \pm 25 \; \rm{J}$ was delivered on target in a full-width-half-maximum (FWHM) pulse length of $\tau_{L} = 1 \pm 0.2 \; \rm{ps}$. It was focused with an $f/3$ off-axis parabolic mirror to a $w_{0} = 10 \pm 2 \; \mu \rm{m}$ FWHM focal spot diameter containing up to $50 \%$ of the laser pulse energy to produce a mean peak vacuum intensity of $(5.3 \pm 1.8) \times 10^{19} \; \rm{Wcm}^{-2}$, corresponding to an $a_{0} \approx 6.5 \pm 2.2$. The prepulse energy was measured using a fast photodiode behind a water-cell to be $16 \pm 5 \; \rm{mJ}$ (measurements available for about $20 \%$ of the shots), giving a nanosecond energy contrast ratio of $\sim 10^4$. The laser pulse was linearly polarized and had a wavelength of $\lambda_{L} = 1.053 \; \mu \rm{m}$, so therefore $n_{c} = 10^{21} \; \rm{cm}^{-3}$.

Very low-density foams were used, with mass densities of $3$--$100 \; \rm{mg/cc} \pm 5\%$ that fully ionize to produce plasma with electron number density range $(0.9$--$30) \times 10^{21} \; \rm{cm}^{-3}$ (previously used for the experiments in Refs.\ \cite{Willingale_PRL_2009, Willingale_PoP_2011}) to produce well-controlled near-critical density targets. The low density foam targets were fabricated using the \textit{in situ} polymerization technique and had a composition of $71 \%$ C, $27 \%$ O and $2 \%$ H by mass. The pore and thread structures were sub-micron, so a relatively homogenous plasma was expected on the $\lambda_{L}$ scale. The delicate foams were supported within $250 \; \mu \rm{m}$ thick washers, with the aperture filled with foam to produce $(250 \pm 20) \; \mu \rm{m}$ thick foam targets. The angle of incidence of the laser pulse onto the front surface of the foam at $s$-polarization was $16^{\circ}$. For comparison, some shots were taken onto Mylar foils (fully ionized plasma density of $433 n_{c}$, i.e.\ $\gg n_{c}$), with thicknesses of $23 \; \mu \rm{m}$, $67.5 \; \mu \rm{m}$ or $250 \; \mu \rm{m}$.

\begin{figure}
\centering
    	\includegraphics[width=0.5\columnwidth]{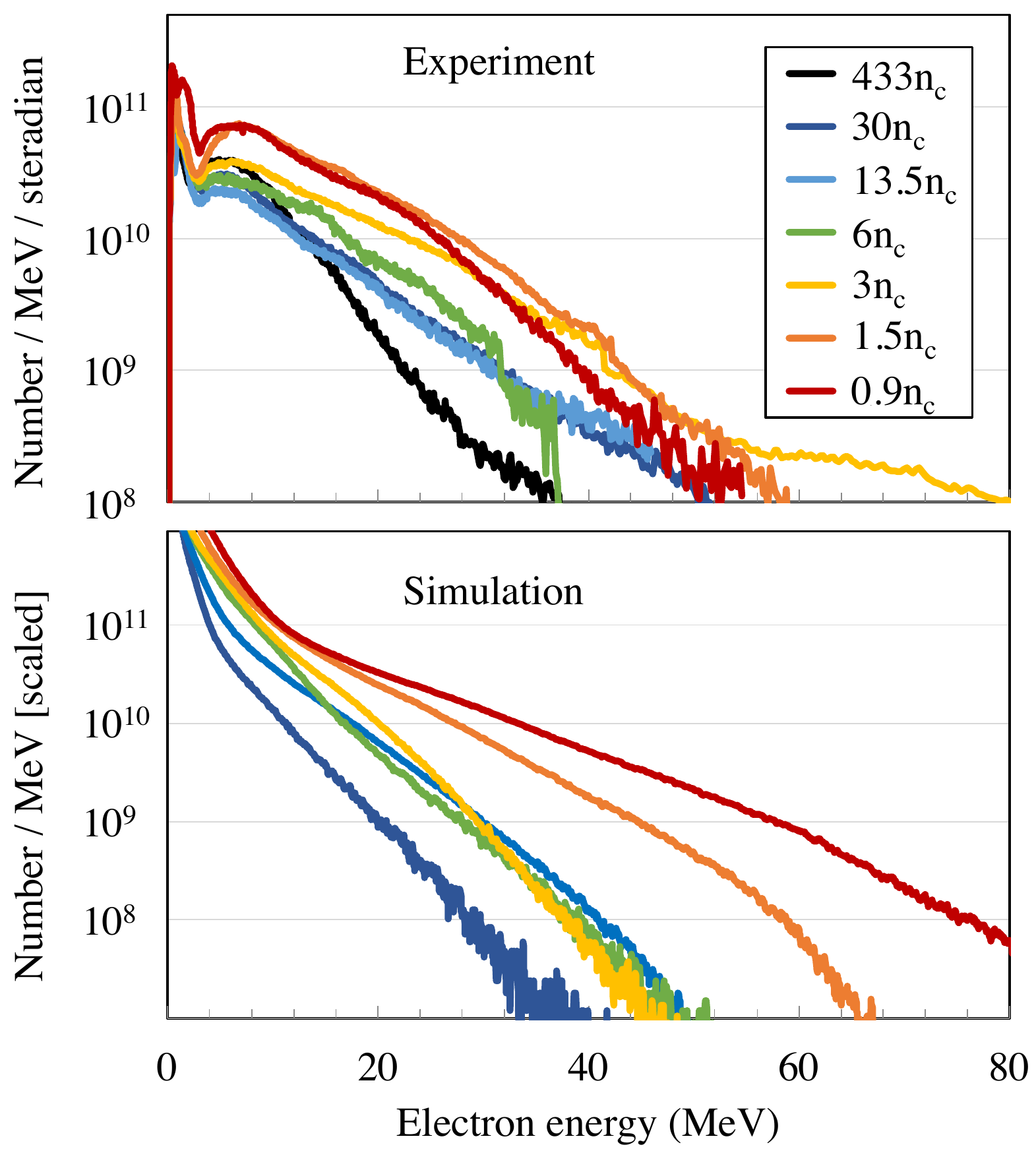}
\caption{Measured spectra, averaged over each density (upper plot) and simulated (lower plot) electron spectra from different density targets. The spectra were measured along the laser-axis direction. The simulated spectra are snapshots for the entire plasma volume.}
\label{Figure_1}
\end{figure}

\begin{figure}
	\centering
    	\includegraphics[width=0.7\columnwidth]{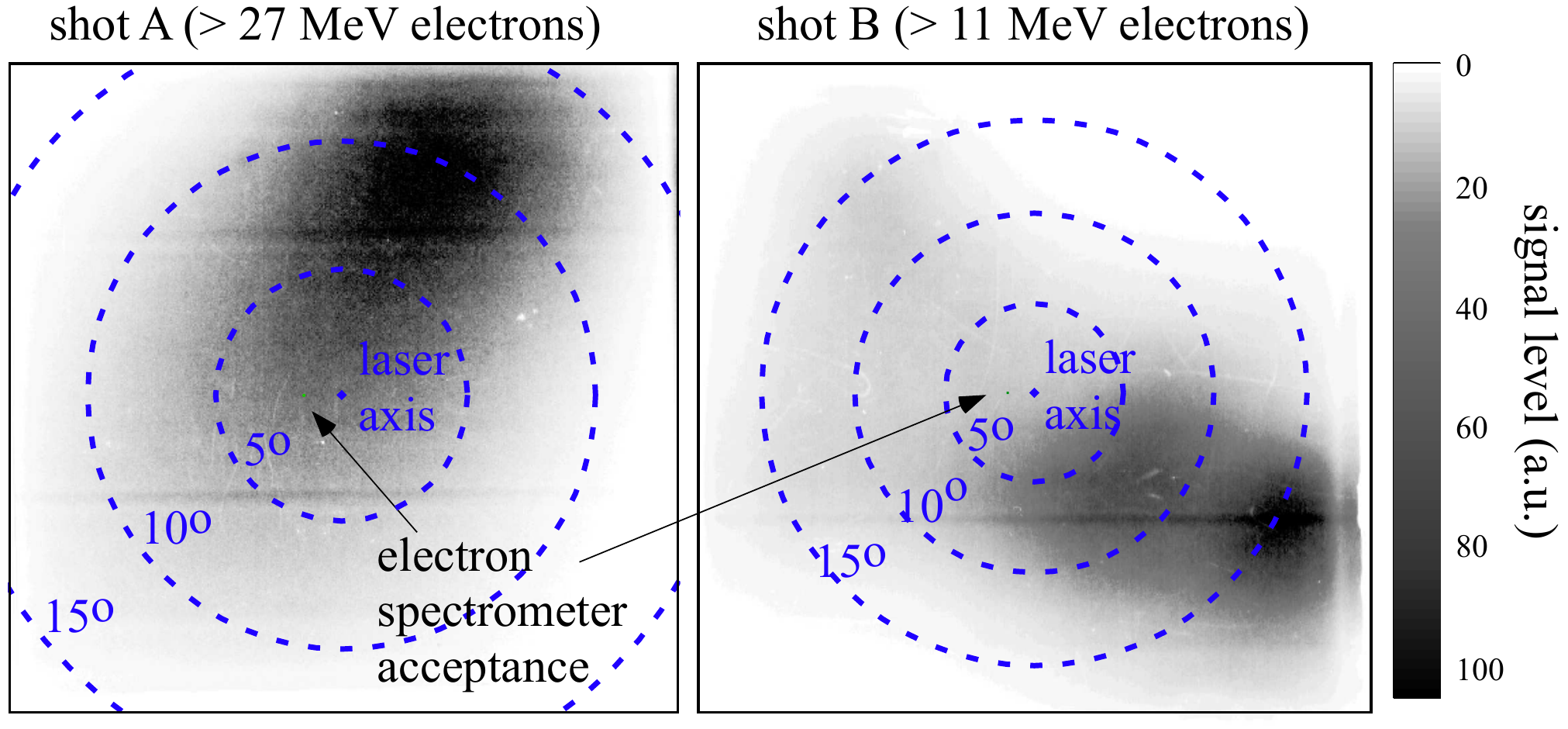}
  \caption{Electron beam divergence and pointing from two different shots onto $1.5 n_{c}$ plasma. The electron spectrometer acceptance angle and position is shown as the orange dot.}
\label{Figure_2}
\end{figure}

\section{Particle-in-cell simulation parameters}
\label{PIC_section}

To gain insight into electron heating in near-critical plasmas, two-dimensional simulations were performed using a fully-relativistic particle-in-cell code EPOCH~\cite{EPOCH} for the same range of near-critical target densities.
The laser propagates along the $x$-axis, and is linearly polarized with the electric field in the $y$-direction.
The laser pulse was approximated by a Gaussian beam focused to a 14 $\mu$m spot (FWHM of intensity) with $\lambda = 1.053 \; \mu \rm{m}$ and 0.7 ps in duration.
The peak vacuum normalized vector potential was $a_0 = 6.5$.
This laser pulse duration was chosen to mimic the experimental setup while keeping the simulation box in the case of lower density targets manageable.
At lower densities, the laser pulse easily propagates through the plasma.
In order to prevent the laser pulse from burning though the target during the simulation, the plasma thickness would have to be increased by roughly $\delta l \approx c \delta t$ if the pulse duration is increased by $\delta t$.
Additionally, the plasma width would have to be increased as well, because instabilities cause unpredictable and sometimes significant changes of direction for the propagating laser pulse.
Again, the lower density runs are much more impacted by this than the runs with intermediate densities.

Initially, the plasma is uniform, with a sharp boundary at $x = 0$.
The cell size in all the runs was 0.02 $\mu$m by 0.04 $\mu$m to resolve the dynamics of the accelerated electrons~\cite{Arefiev_PoP_2015}.
There were 100 macro-particles per cell at $n_e = 30 n_c$ and $n_e = 13.5 n_c$, and 50 macro-particles per cell in the other runs.
The ratio of macro-particles in each cell representing electrons, protons, carbon ions, and oxygen ions was set at 10:2:7:1.
No ionization took place during the simulation, with the ionization states for carbon and oxygen ions set at $Z_C = 6$ and $Z_O = 8$.
To ensure that the plasma is initially quasineutral, the ion densities are initially set at $n_p = 0.04 n_e$ for protons, $n_C = 0.116 n_e$ for carbon ions, and $n_O = 0.033 n_e$ for oxygen ions, so that $n_p + Z_C n_C + Z_O n_O = n_e$.
The target thickness in each case was sufficient to prevent the laser pulse from burning through the target during the runs that lasted 2 ps for $n_e = 30 n_c$ and $n_e = 13.5 n_c$ and 2.5 ps for $n_e = 0.9 n_c$, $n_e = 1.5 n_c$, $n_e = 3 n_c$, and $n_e = 6 n_c$.
Specifically, the target thickness was 140 $\mu$m for $n_e = 0.9 n_c$, 110 $\mu$m for $n_e = 1.5 n_c$, 60 $\mu$m for $n_e = 3 n_c$ and $n_e = 6 n_c$, and 25 $\mu$m for $n_e = 13.5 n_c$ and $n_e = 30 n_c$.
Using shorter targets made these computationally demanding runs more manageable, particularly in the case of high density targets where the number of macro-particles per cell had to be doubled.


\section{Results}
\label{results_section}

\subsection*{Experimental results}

The experimental electron spectra were measured using magnetic electron spectrometers \cite{Chen_RSI_2008} with image plate detectors. The upper plot in Fig.~\ref{Figure_1} shows typical electron spectra measured along the laser axis for each target density.
The lower plot in Fig.~\ref{Figure_1} shows snapshots of the simulated electron spectra at the peak of the laser intensity.
The maximum vacuum transverse and longitudinal electron $\gamma$ associated with $a_0 = 6.5$ are $p_{y}/m_e c = a_0 = 6.5$ and $p_{x}/m_e c = a_0^2 / 2 = 22$ respectively, so $n_{\gamma c}$ is likely in the range $6.5n_c$--$22n_c$.
Both plots show higher maximum electron energies for near-critical target densities when compared with relativistically opaque densities, i.e.\ $30 n_{c}$.
The experimental data shows significant fluctuations at the lowest electron densities.
The likely explanation for this is a variable electron beam pointing, as illustrated in Fig.~\ref{Figure_2}.
The electron beam divergence, $\theta_{e}$, and pointing were measured using a stack of aluminum and image plate layers.
Figure~\ref{Figure_2} shows electrons beams from two different $n_{e} = 1.5 n_{c}$ shots with $\theta_{e} \leqslant 10^{\circ}$ (half angle). The beams have asymmetric distributions and shot B has hints of more than one beam.
These measurements also indicate that the electron beam pointing was unstable.
The center of the beam was offset by $> 10^{\circ}$ from the original laser-axis with apparently arbitrary and random direction.

These observations are consistent with the numerical modeling where for $n_e =  0.9 n_c$ and $1.5 n_{c}$ the simulations showed unstable beam propagation accompanied by significant off axis deviations.
The total electron spectra from the simulation should be unaffected by this instability, but it could lead to a seeming decrease of the measured electron spectrum at $n_e =  0.9 n_c$.
The feature of primary interest to us here is that the spectra from the relativistically near-critical target range of $3 n_c$ to 13.5$n_c$ exhibit a similar looking energetic electron tail.
The spectrum drops only as the density is increased to $n_e =  30 n_c$ and the target becomes relativistically opaque and hence overdense.
These trends are in agreement with the simulation study presented in reference \cite{Willingale_PoP_2011}, where the simulated electron spectra from different near-critical density targets are considered, but the electron acceleration mechanisms were not investigated.

\begin{figure}
	\centering
    	\includegraphics[width=0.9\columnwidth]{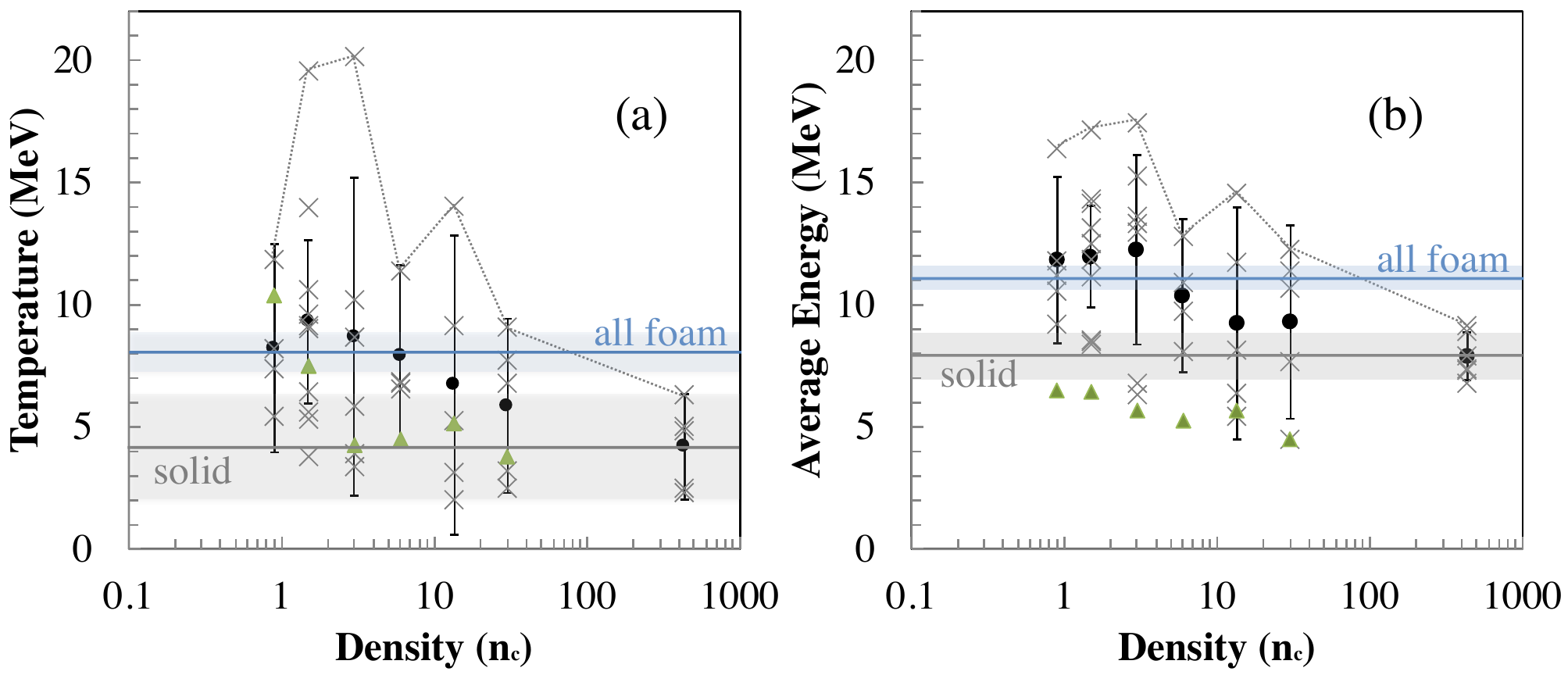}
  \caption{The experimental $T_{e}$, (a), and average electron energy, (b), extracted from the spectra along the laser-axis direction. The crosses show the individual shot data, whereas the circles give the averaged data for each density with the corresponding error-bars showing the $95\%$ confidence interval using Student's t-distribution. The black line shows the average solid target values and shaded region the error and the blue line shows the average with the standard error (shaded region) over the foam target shots and to guide the eye, the dashed line shows the maximum values at each density. The error on the individual shots is not shown for clarity, but typical errors are $\sim 10 \%$. The green triangles show the simulation temperature and average energy.}
\label{Figure_3}
\end{figure}

The experimental spectra were generally reasonably exponential so a fit was made to the data to determine a Maxwellian-like temperature, $T_{e}$, along the laser axis and are plotted versus plasma density in Fig.~\ref{Figure_3} (a), albeit with significant error in some cases.
Individual shot data is plotted as crosses and the mean for each density is plotted by circles with the error-bars showing the $95\%$ confidence interval using Student's t-distribution.
The likely reasons for the fairly large variation in $T_{e}$ are the variable electron beam pointing, as already discussed, and uncertainties when fitting to data with non-Maxwellian features.

The average electron energy measured between $2 \; \rm{MeV}$ and the detection threshold is a different way to present the data (Fig.~\ref{Figure_3} (b)).
There was smaller shot-to-shot variation for the average electron energies making the trend clearer and the mean values (squares) have a reduced standard deviation.
For the highest density, the solid Mylar foil targets ($n_{e} = 433 n_{c}$), and $n_e = 30n_c$ foam, the $T_{e}$ is in reasonable agreement with $U_{p} \approx 5.4 \; \rm{MeV}$ for $a_{0} = 6.5$.
For the lower densities, the high-energy tail enhances the $T_{e}$ and average electron energy to significantly above $U_{p}$.
For both $T_e$ and the average energy, the solid target mean values (gray lines) are significantly lower that the mean values over all of the foam target shots (blue lines).

Also shown in Fig.~\ref{Figure_3} as green triangles are the $T_e$ extracted from the simulation spectra.
The trends in both $T_e$ and average electron energy are similar, albeit with slightly lower values.
This shift is likely due to the difference between two- and three-dimensional effects, as well as the larger effective collection angle for calculating the simulation spectra.

\begin{figure}
\centering
\includegraphics[width=0.6\columnwidth]{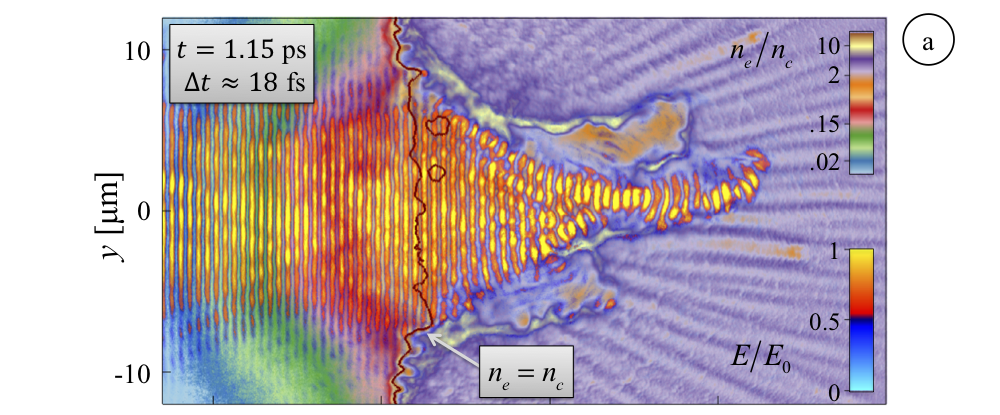}
\includegraphics[width=0.6\columnwidth]{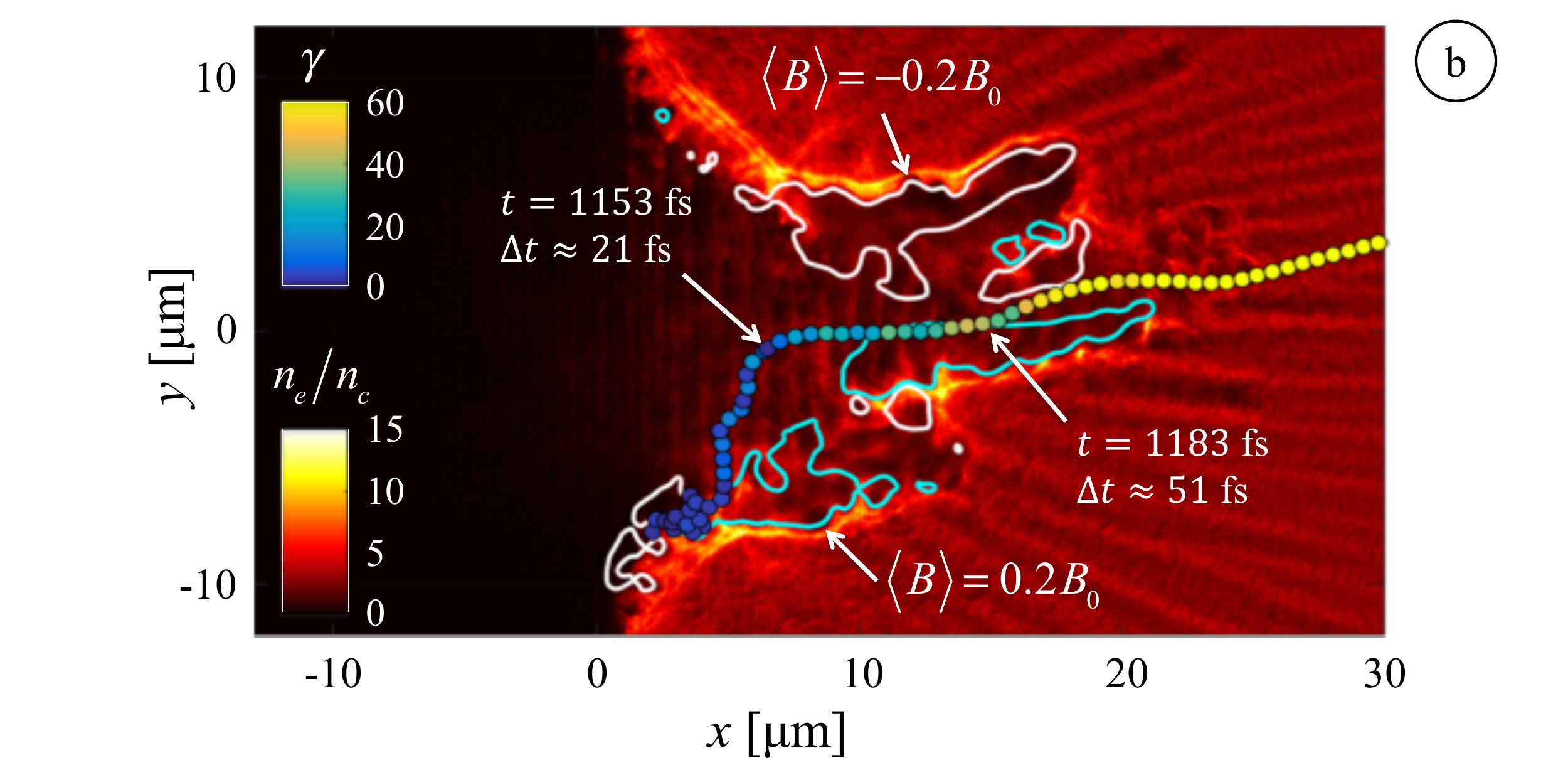}
\caption{Data from the $n_e = 3 n_c$ simulation at $\Delta t \approx 18 \; \rm{fs}$ after the peak of the laser pulse has arrived at $x =0$ $\mu$m (elapsed time since the beginning of the simulation is $t = 1.15 \; \rm{ps}$). (a): $n_{e}$ on a logarithmic scale and the $n_{e} = n_{c}$ contour is indicated. The total electric fields normalized to the peak electric field in the absence of the target, $E_{0}$, is overlaid to highlight the relativistically transparent channel. (b): the same $n_{e}$ on a linear scale with quasi-static magnetic field contours shown. Overlaid is an example electron trajectory that is color-coded to indicate the $\gamma$ at each position.}
\label{Figure_4}
\end{figure}

\subsection*{Simulated electron energy gain}

The key features of the laser-plasma interaction in the near-critical regime ($n_c < n_e < n_{\gamma c}$) observed in the PIC simulations are illustrated in Fig.~\ref{Figure_4}.
The electron density prior to the interaction with the laser pulse is uniform, with $n_e = 3 n_c$.
The intense laser pulse induces relativistic transparency, which allows it to propagate through the plasma beyond the $n_e = n_c$ surface shown with a red curve in Fig.~\ref{Figure_4}a.
The electric field amplitude $E$ in Fig.~\ref{Figure_4}a has distinct spatial modulations associated with the oscillating field of the laser pulse more than 20 $\mu$m beyond the $n_e = n_c$ surface.
The density and the field snapshots are taken at $\Delta t \approx 18$ fs after the peak intensity would have arrived at $x = 0$ $\mu$m in the absence of the plasma.
The elapsed time since the beginning of the simulation is $t = 1.15$ ps.

The laser pulse produces a narrowing, funnel-like channel in the plasma with a laser-driven longitudinal electron current that generates and sustains a relatively strong slowly evolving magnetic field $B_z$.
$B_z$ is averaged over ten laser periods to find the quasi-static component that denoted as $\langle B \rangle$.
Two contours, $\langle B \rangle = \pm B_0$, are shown in Fig.~\ref{Figure_4}b, where $B_0$ is the peak amplitude of the laser magnetic field in the absence of the plasma.
Evidently, the quasi-static magnetic field is not negligible compared to the magnetic field of the laser and should be expected to impact the electron dynamics inside the funnel-like channel~\cite{Stark_PRL_2016, Jansen_PPCF_2018}. 

\begin{figure}
\centering
\includegraphics[height=8cm]{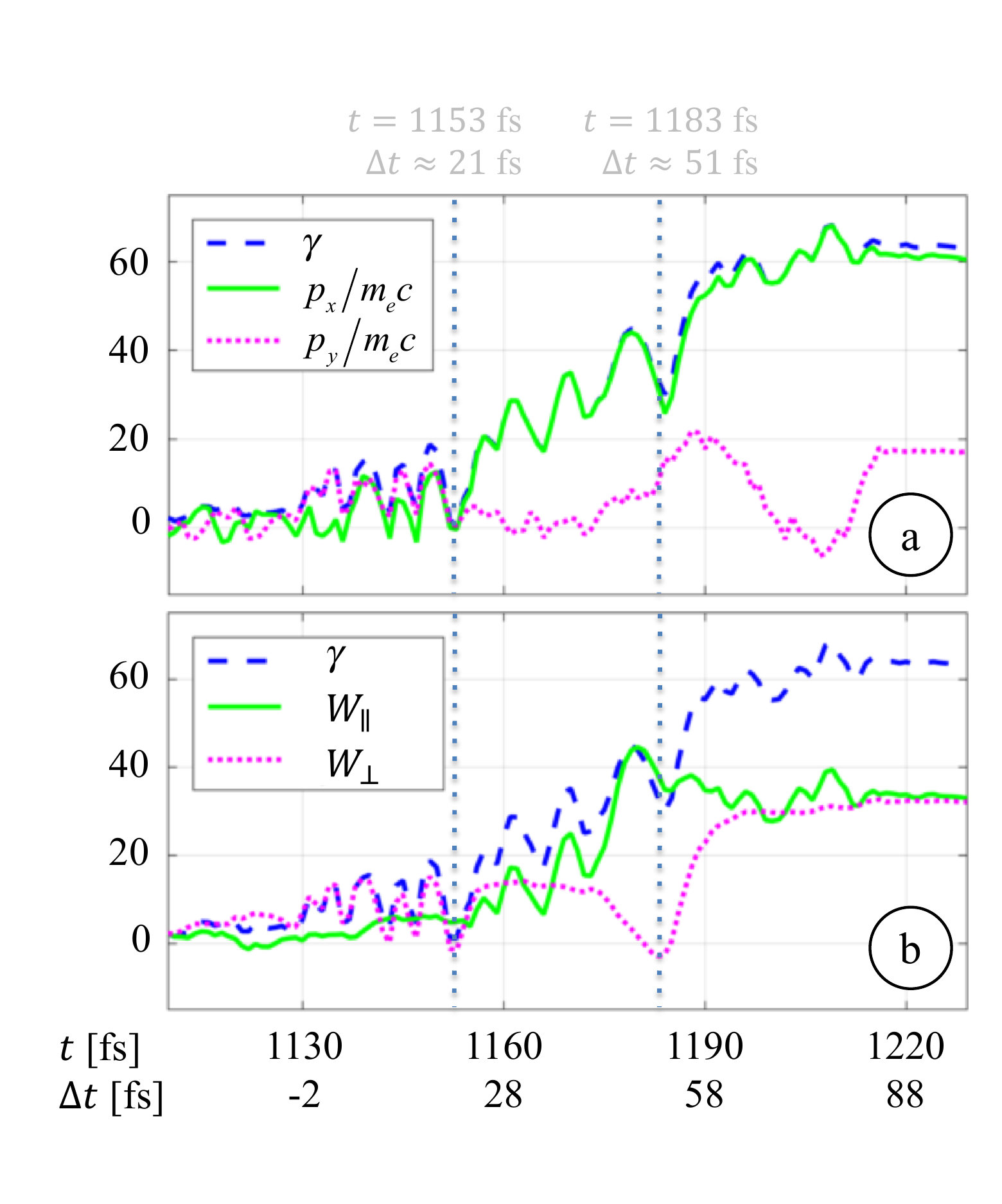}\includegraphics[height=8cm]{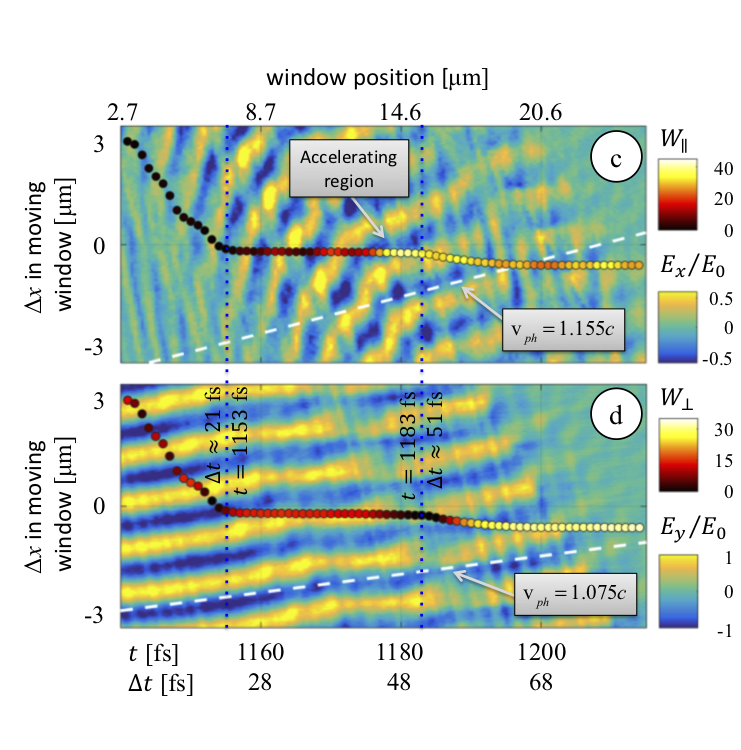}
\caption{(a-b): The example electron $\gamma$ (blue dashed) as a function of time. The longitudinal (green line) and transverse (pink dots) components of the electron momentum (a) and contributions to the energy gain due to the each electric field component (b) are shown. (c-d): The longitudinal (c) and transverse (d) electric fields in a window moving along the $x$-axis with $c$. The location of the center of the window is shown above panels as a function of the elapsed time $t$ since the beginning of the simulation. The example relative electron position is color-coded according to the energy gain ($\gamma$) from the corresponding electric field component.}
\label{Figure_5}
\end{figure}

The energetic electrons are tracked during their energy gain process and the majority of the electrons from the energetic tail are found to originate inside this relativistically transparent channel.
Figure~\ref{Figure_4}b shows a representative electron trajectory to be discussed in detail.
As evident from the color-coded $\gamma$-factor in Fig.~\ref{Figure_4}b, the energy gain for this electron takes place well inside the plasma where $n_e > n_c$.
Figure~\ref{Figure_5}a shows the time evolution of the electron momentum components and the $\gamma$-factor for the same electron, illustrating that the electron is accelerated primarily in the laser propagation ($x$) direction.
To determine the underlying mechanism, the contributions to the $\gamma$-factor from the work done by the transverse, $E_y$, and longitudinal, $E_x$, components of the electric field are calculated and shown in Fig.~\ref{Figure_5}b as functions of time.
Here we use the following definitions:
\begin{eqnarray}
&& W_{\parallel} \equiv - \frac{1}{m_e c^2} \int |e| E_{x} v_{x} dt,  \label{Wx}\\
&& W_{\perp} \equiv - \frac{1}{m_e c^2} \int |e| E_{y} v_{y} dt, \label{Wy}
\end{eqnarray}
so that $W_{\parallel} + W_{\perp} = \gamma -1$.
Remarkably, half of the energy gained by this tracked electron is contributed by $E_x$. 

The significant role of the longitudinal field is unexpected, since the longitudinal component is negligible in the considered incoming beam due to the large beam width.
In the incoming beam, it can be estimated from the condition $\nabla \cdot {\bf{E}} = 0$, which yields $|E_x| \approx |E_y| \lambda / R$, where $R$ is the characteristic transverse scale of $E_y$.
Taking into account that $R \approx \sqrt{2} w_0$, we find that $|E_x| \approx 0.05 |E_y| \leq 0.05 E_0$, where $E_0$ is the amplitude of the transverse electric field in the focal plane of the incoming laser pulse.
In order to determine the actual fields experienced by the considered electron as it travels into the target, we use a window that is moving with the speed of light along the beam axis ($x$-axis).
The tracked electron is in the center of the window when it begins its longitudinal motion at $t = 1153$ fs (from the beginning of the simulation) and $x = 6.55$ $\mu$m.
Figures~\ref{Figure_5}c and \ref{Figure_5}d show $E_x$, $E_y$, and the longitudinal electron displacement in the moving window.
In contrast with the transverse field, a strong longitudinal electric field with $|E_x| \sim0.5 E_0$ emerges well inside the near-critical plasma ($x > 6.55$ $\mu$m).
This is the field that contributes to the electron energy gain, rather than the weak longitudinal field that we estimated for the incoming beam before it enters the target.
The mechanism responsible for generating this field is explained towards the end of this section, but here we simply point out that it is critical for the electron acceleration: the simulations observe a 30 fold increase in the longitudinal field compared with the vacuum case.

\begin{figure}
\centering
\includegraphics[width=1.1\columnwidth]{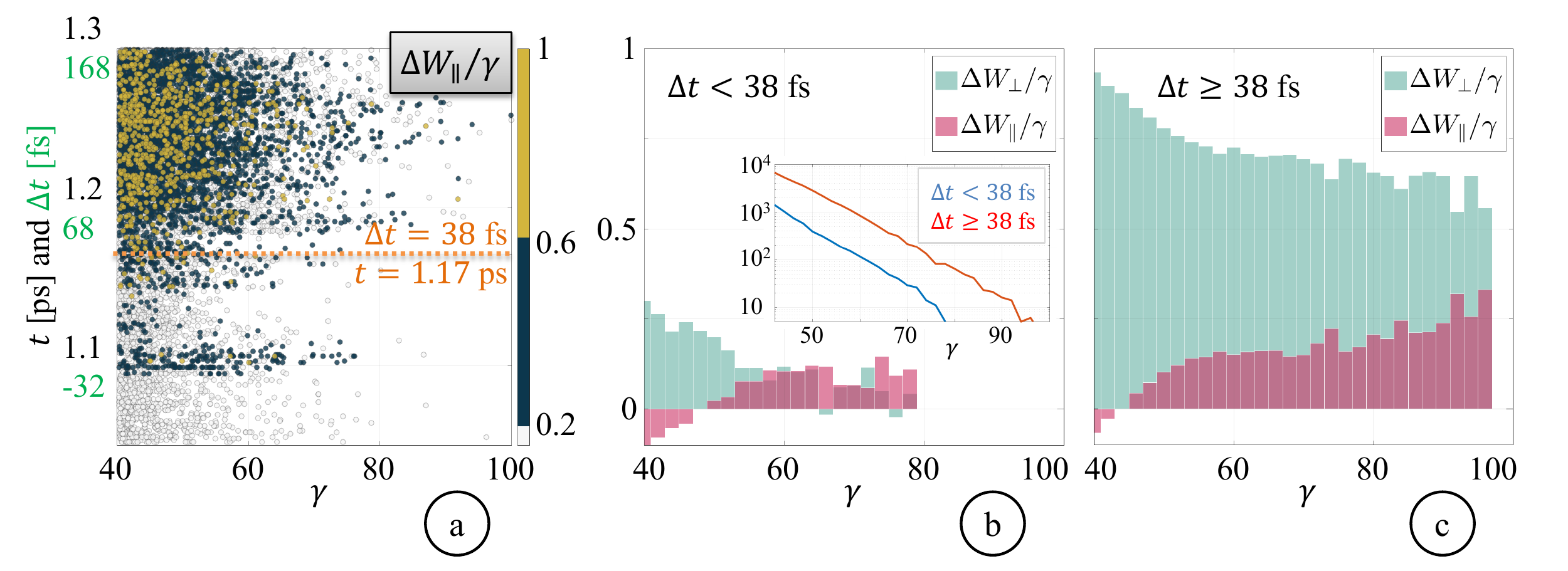}
\caption{Electron heating in the $n_e = 3 n_c$ simulation during the time interval of 1.05 ps $ \leq t \leq$ 1.3 ps. The electrons are tracked inside a box with $|y| < 8$ $\mu$m and $x < 30$ $\mu$m during 1050 fs $< t < $ 1300 fs. The panels show the electron data for the electrons that leave the box with $\gamma > 40$ moving to the right through the boundary located at $x = 30$ $\mu$m during 1050 fs $< t < $ 1300 fs. (a) shows a relative contribution, $\Delta W_{\parallel}/\gamma$, of the work done by the longitudinal field towards the total energy of each tracked electron. (b) and (c) show a statistical analysis of the components of the work done for $t < 1.17 ps$ ($\Delta t < 38 \; \rm{fs}$) and $t \geqslant 1.17 ps$ ($\Delta t \geqslant 38 \; \rm{fs}$) respectively. The inset shows the count of macro-particles representing electrons in panels (b) and (c).}
\label{Figure_6}
\end{figure}

The electron momentum is primarily longitudinal and, in agreement with Eq.~(\ref{Wx}), this enables a rapid transfer of energy from $E_x$ to the electron, shown with the color-coded circles in Fig.~\ref{Figure_5}c.
The electron gained the remainder of its energy from the transverse field where the self-generated magnetic field plays an important role in enabling this energy transfer.
The initial contribution right after the electron reaches the axis of the beam and begins its longitudinal motion (see Fig.~\ref{Figure_4}) is made via the conventional direct laser acceleration mechanism.
However, the presence of the near-critical plasma considerably limits the resulting energy gain by increasing the wave phase velocity $v_{ph}$ and thus deteriorating the phase matching.
As shown in Fig.~\ref{Figure_5}, the phase velocity of the transverse electric field in side the channel is $v_{ph} \approx 1.075 c$.
According to Ref.~\cite{Robinson_PoP_2015}, we should expect an energy gain corresponding to $\gamma \approx a_0 \left[ 2 (v_{ph} - c)/c \right]^{-1/2} \approx 16$.
This matches well the $E_y$-contribution at about 1160 fs shown in Fig.~\ref{Figure_5}b.
The second significant increase in $W_{\perp}$ occurs after the electron encounters a region with a strong magnetic field at 1183 fs and becomes deflected (see Fig.~\ref{Figure_4}).
The transverse momentum increases as a result of the deflection, which is typically detrimental for the direct-laser-acceleration.
The magnetic field however also breaks the synchronism between $p_y$ and $E_y$ that otherwise prevents further energy gain.
Following the deflection, the electron enters a region of negative $E_y$ (see Fig.~\ref{Figure_5}d) with a substantial positive transverse momentum $p_y$ (see Fig.~\ref{Figure_5}a).
This then allows for a rapid transfer of energy shown in Fig.~\ref{Figure_5}d with the color-coded circles, similar to what was observed in the case of $E_x$. 

Detailed electron tracking has also enabled us to determine average relative contributions by $E_x$ and $E_y$ over a wide range of electron energies, shown in Fig.~\ref{Figure_6}.
We have tracked electrons in a box enclosing the funnel-like channel, $|y| < 8$ $\mu$m and $x < 30$ $\mu$m, recording $W_{\parallel}$ and $W_{\perp}$ over 250 fs (1050 fs $< t < $ 1300 fs).
We show the results for electrons with $\gamma > 40$ that leave the box moving to the right through the boundary located at $x = 30$ $\mu$m during 1050 fs $< t < $ 1300 fs.
Figure~\ref{Figure_6}a shows a relative contribution, $\Delta W_{\parallel}/\gamma$, of the work done by the longitudinal field towards the total energy of each tracked electron.
As the funnel structure becomes more pronounced with time, the effect of the longitudinal electric field becomes more pronounced.
After $t \approx 1.17$ ps, there are electrons, shown with yellow markers, that have gained more than 60\% of their total energy from $E_x$. 

The energy exchange with $E_x$ is positive only for some electrons, while others lose an appreciable amount of energy to $E_x$.
Figures~\ref{Figure_6}b and \ref{Figure_6}c provide a statistical analysis of the electron heating in order to determine the effect of $E_x$ for each energy range.
We split the electrons into those that leave the box before and after $t \approx 1.17$ ps. 
For the electrons that leave at $t <1.17$ ps, most of the energy had been accumulated outside of the spatial region of interest or before we started tracking them.
For the electrons that leave after $t  \approx1.17$ ps, most of the energy is accumulated inside the region with the funnel-like channel.
The inset in Fig.~\ref{Figure_6}b shows the count of the macro-particles representing electrons in the histograms of Figs.~\ref{Figure_6}b and \ref{Figure_6}c.
The curves are essentially the electron spectra.
They confirm that the heating for the first group is ineffective, so its contribution compared to that of the second group is relatively insignificant.
The most important trend for the second group is that the longitudinal electric field contributes a considerable amount of energy of the energetic electrons, with $\Delta W_{\parallel} / \Delta W_{\perp} \approx 0.3$ for $\gamma > 80$.
Contrary to what one might expect, the work by the transverse electric field inside the region of interest never exceeds 70\% of the total energy for the energetic electron tail with $\gamma > 60$.

\subsection*{Accelerating field structure}

We have determined that the longitudinal electric field that arises inside the narrowing plasma channel makes an appreciable contribution towards the electron energy gain.
Here, we show that it is caused by reflections of the incoming laser beam off the walls of the funnel-like channel rather than by beam focusing or space-charge effects.

\begin{figure}
\centering
\includegraphics[width=1.0\columnwidth]{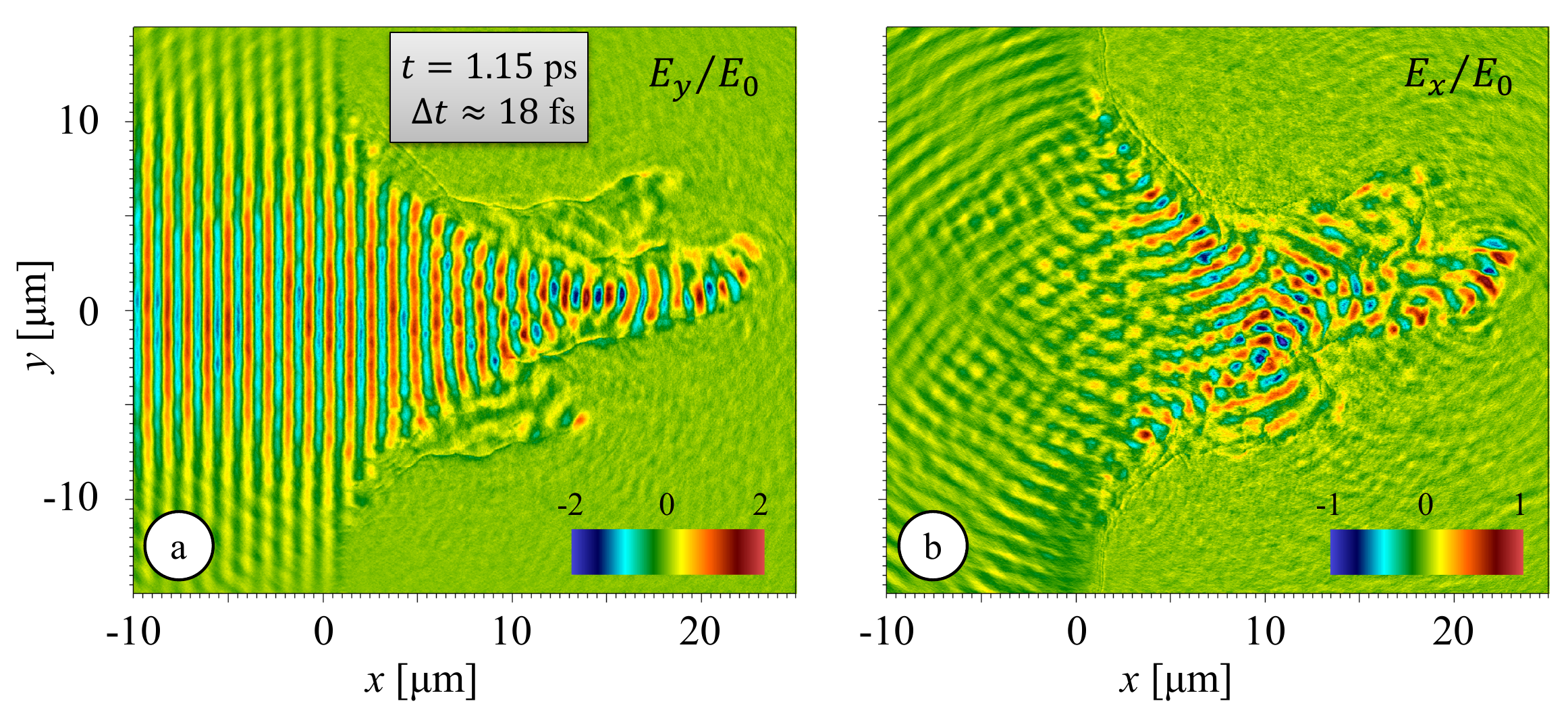}
\caption{The transverse and longitudinal electric fields shown at the same time as the images in Fig.~\ref{Figure_4}  ($t = 1.15$ fs and $\Delta t \approx 18$ fs). Both components are normalized to $E_0 \approx 2 \times 10^{13}$ volt/m, the peak amplitude of the electric field in the incoming laser beam in the absence of the plasma. The maximum and minimum values of these field components are: $\max \left( E_y/E_0 \right) \approx 1.8$,  $\min \left( E_y/E_0 \right) \approx -1.7$,  $\max \left( E_x/E_0 \right) \approx 0.9$, and  $\min \left( E_y/E_0 \right) \approx -1.1$.}
\label{Figure_7}
\end{figure}

Snapshots of $E_y$ and $E_x$ shown in Fig.~\ref{Figure_7} have seemingly uncorrelated patterns.
The transverse component $E_y$ has almost flat wave-fronts as deep as 10 $\mu$m into the plasma.
In contrast to that, $E_x$ has what appears as tilted wave-fronts, such as in the region with $y>0$ $\mu$m and 0 $\mu$m $< x < 10$ $\mu$m where the wave-fronts of $E_y$ are still flat.
In the case of beam focusing, the wave-fronts of $E_x$ and $E_y$ are aligned (for example, see Ref.~\cite{Gong_arXiv_2018} where a narrow channel is used to amplify $E_x$).
However, this pattern is not visible in the incoming beam because the corresponding field, $|E_x| \approx 0.05 E_0$, is too weak.
The focusing in the narrowing channel is also insufficient to explain the observed increase of the longitudinal field.
The beam width would have to decrease at least by a factor of ten for $E_x$ to be visible in Fig.~\ref{Figure_7}b, but the beam width decreases by not more than a factor of two when $E_x$ becomes strong. 

\begin{figure}
	\centering
    	\includegraphics[width=0.9\columnwidth]{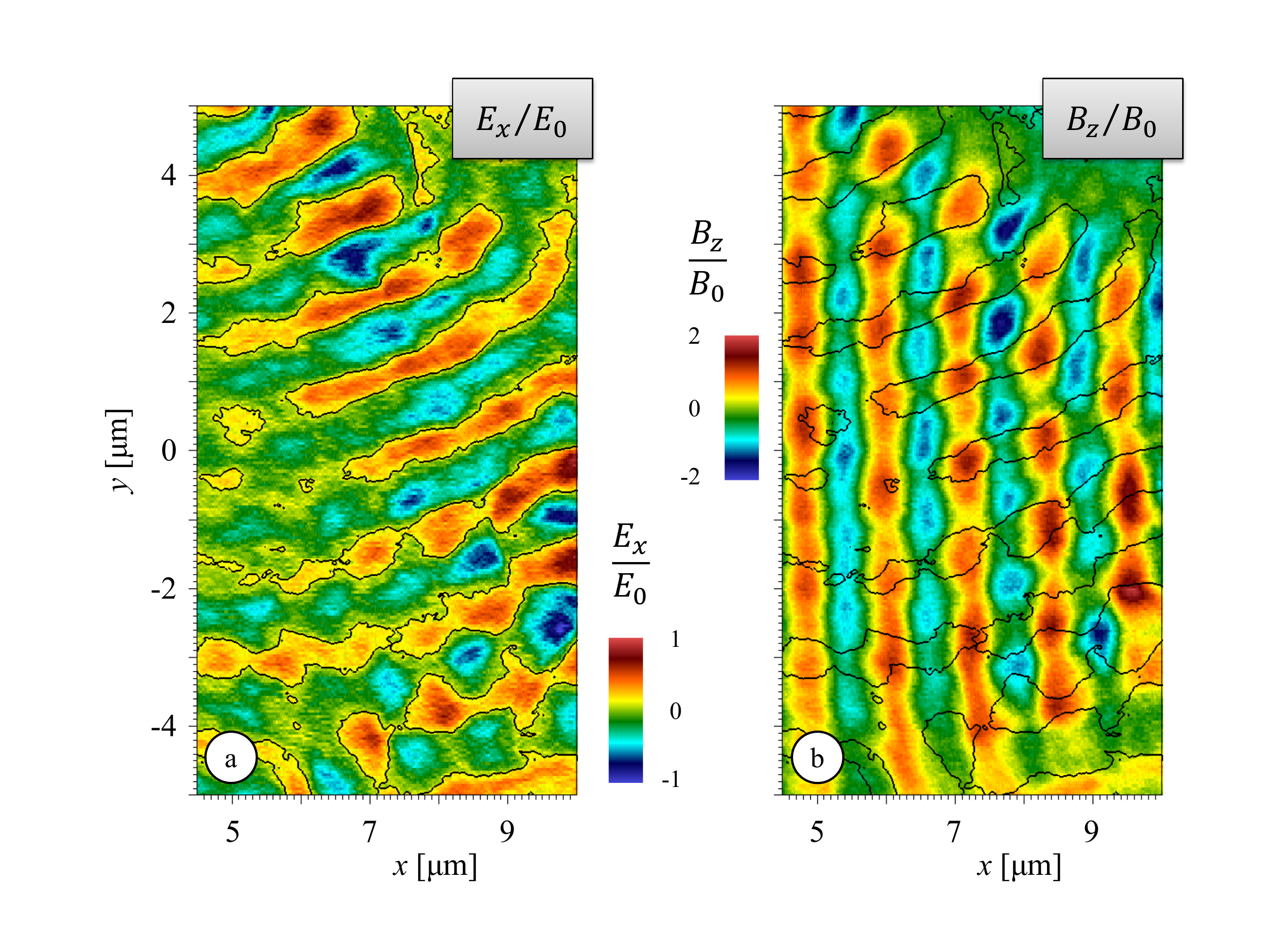}
  \caption{Magnified region of the longitudinal electric field $E_x$ (a) and transverse magnetic field $B_z$ (b) at  $t = 1.15$ fs. The magnetic field is normalized to $B_0 \approx 66.2$ kT, which is the peak amplitude of the magnetic field in the incoming laser beam in the absence of the plasma. The black curves in both panels indicate the contours of constant $E_x$, with $E_x / E_0 = 0.1$. Over the entire simulation domain, $\max \left( B_z/B_0 \right) \approx 2.1$ and $\min \left( B_z/B_0 \right) \approx -1.8$.}
\label{Figure_8}
\end{figure}

Figure~\ref{Figure_8} shows magnified snapshots of $E_x$ and $B_z$ in the region with tilted wave-fronts of the longitudinal electric field.
A comparison of Figs.~\ref{Figure_7}a and \ref{Figure_7}b reveals that the transverse periodic modulations of $B_z$ coincide with the wave-fronts of $E_x$ that are shown with contours in both panels to guide the eye.
The fact that there is a correlation between $E_x$ and $B_z$ indicates that space-charge effects are unlikely to be the cause of the strong longitudinal electric field.
The modulations are consistent with reflections.

In order to demonstrate the role of beam reflections in creating the observed field structure, we consider a simple model where three plane waves overlap, producing an interference pattern.
The electric and magnetic fields in each of the waves are given by
\begin{eqnarray}
&& E_x = - E_* \sin \theta \cos \left[ 2 \pi x' / \lambda + \psi(t) \right], \\
&& E_y = E_* \cos \theta \cos \left[ 2 \pi x' / \lambda + \psi(t) \right], \\
&& B_z = E_* \cos \left[ 2 \pi x' / \lambda + \psi(t) \right], 
\end{eqnarray}
where $E_*$ is the wave amplitude, $\theta$ is the angle between the $x$-axis and the direction of the wave propagation, $\psi(t)$ is the time-dependent phase and 
\begin{equation}
x' \equiv x \cos(\theta) + y \sin(\theta)
\end{equation}
is the distance along the direction of the wave propagation.
We mimic the case observed in the simulation by assuming that the main wave propagates forward along the $x$-axis, such that $E_* = E_0$, $\theta = 0$.
Without any loss of generality, consider two lower amplitude waves that come in at an angle, where $E_* = 0.25 E_0$, $\theta = - \pi/3$ and $E_* = 0.1 E_0$, $\theta = \pi / 8$.

\begin{figure}
	\centering
    	\includegraphics[width=1.0\columnwidth]{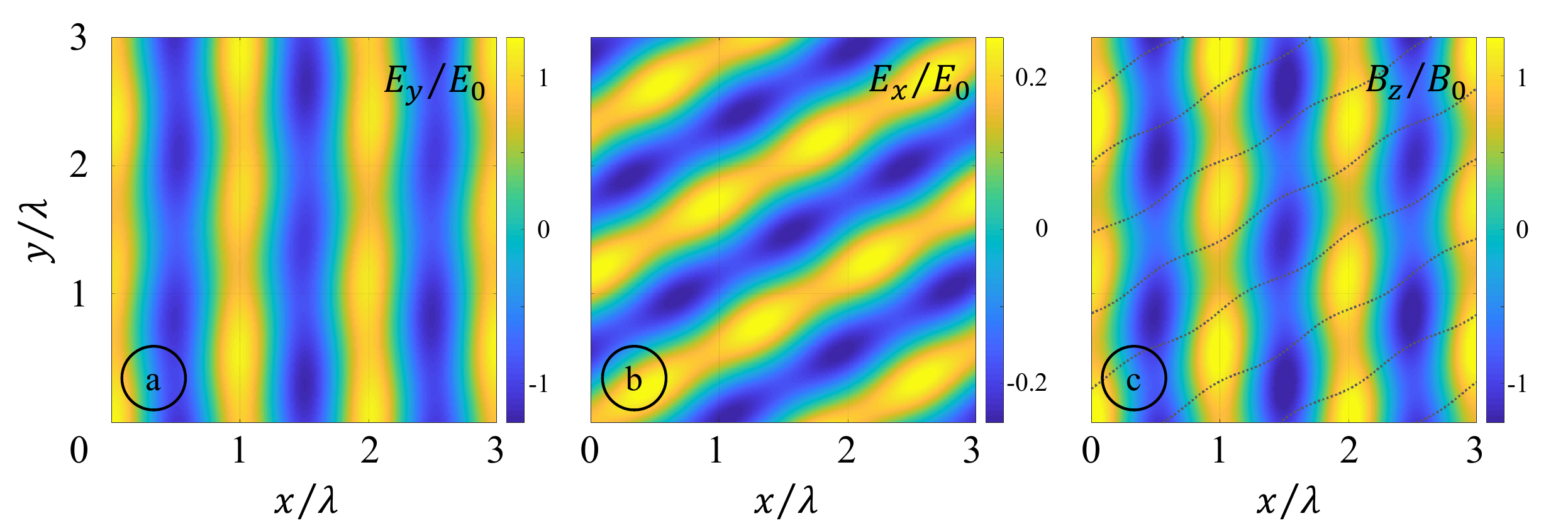}
  \caption{A wave pattern produced by three overlapping plane waves of different amplitude. The dotted curves in the right panel indicate the contours of constant $E_x$, with $E_x / E_0 = 0$.}
\label{Figure_9}
\end{figure}

The interference patterns at $\psi = 0$ for the electric and magnetic fields are shown in Fig.~\ref{Figure_9}.
Similarly to what is seen in Fig.~\ref{Figure_7}, the wave-fronts of $E_y$ are vertical, but the wave-fronts of $E_x$ are clearly tilted without any correlation between the two patterns.
This pattern has a clear origin: the wave-fronts of $E_x$ are created exclusively by the lower-amplitude waves.
The explanation is further corroborated by the difference in the longitudinal phase velocities of $E_x$ and $E_y$ in Figs.~\ref{Figure_5}c and \ref{Figure_5}d.
These results were obtained from the PIC simulation and they show that the wave-fronts of $E_x$ are moving faster.
Since the lower-amplitude waves that are responsible for $E_x$ are moving at an angle with respect to the $x$-axis, their phase velocity along the $x$-axis is indeed increased.

The last point to emphasize is the correlation between the modulation of $B_z$ and the tilted wave-fronts of $E_x$ in Figs.~\ref{Figure_9}b and \ref{Figure_9}c. This pattern is again similar to what is seen in the PIC simulations and shown in Fig.~\ref{Figure_7}. The incoming beam has only one component of the magnetic field, which is $B_z$. Reflections do not alter the polarization of the magnetic field, as opposed to what happens to the electric field. As a consequence, the tilted wave-fronts contribute more to the magnetic field of the main wave than to $E_y$ and that is why the modulations in the magnetic field are much more pronounced than those in the transverse electric field.

This simple model elucidates the mechanism responsible for the observed 30 fold increase in the longitudinal field compared with the vacuum case.
The increase takes place without any significant laser beam focusing.
The field is particularly beneficial for energizing electrons that are accelerated in the forward direction by the pulse, i.e. the main component of the wave.
Since the electron momentum is primarily longitudinal, a rapid transfer of energy from $E_x$ to the electron takes place, shown with the color-coded circles in Fig.~\ref{Figure_5}c.

\section{Summary}
\label{summary_section}

In conclusion, we have shown that laser beam propagation in near-critical plasmas, where $n_c < n_{e} < n_{\gamma c}$, can create conditions favorable for electron heating to energies well beyond what is achievable using transverse electric field DLA in such plasmas. Oscillating longitudinal electric and quasi-static magnetic fields generated by the narrowing plasma channel play a profound role in electron heating, enabling rapid and significant energy transfer to electrons from the laser pulse despite the appreciable super-luminal phase velocity. On average, the longitudinal electric field contributes roughly one third of the energy transferred by transverse electric field of the laser pulse to electrons of the super-ponderomotive tail.

Situations where this mechanism may be particularly important are in thin foil targets that decompress to near-critical densities on the timescale of the laser pulse \cite{Palaniyappan_NP_2008, Howell_NJP_2015}, for neutron beam generation \cite{Pomeratz_PRL_2014}, for hole-boring fast ignition \cite{Tabak_PoP_1994}, or for the next generation of laser systems, currently under construction, that will reach intensities accessing a ``QED-plasma'' regime -- where non-linear synchrotron $\gamma$-ray production and multi-photon Breit-Wheeler pair production become important -- and even solid aluminum targets will be in the $n_{\gamma c}$ regime \cite{Ridgers_PRL_2012}.

\ack
$^{\ddag}$Corresponding author.
$^{\dag}$Present address: Department of Physics and Astronomy, University of California, Irvine.
$^{\ast}$Present address: General Atomics, CA, USA.
$^{\ast \ast}$Present address: Naval Research Laboratory.
The authors gratefully acknowledge technical assistance from the staff of the Jupiter Laser Facility.
Work done by Lawrence Livermore National Laboratory was supported by the U.S. Department of Energy under Contract No. DE-AC52-07NA27344.
LW acknowledges support from the Department of Energy National Nuclear Security Administration under Award Number DE-NA0002028 and DE-NA0002723. AVA was supported by the U.S. DoE through agreements No. DE-NA0002008 and by the National Science Foundation (Grant No. 1632777). Simulations were performed using EPOCH code (developed under UK EPSRC Grants No. EP/G054940/1, No. EP/G055165/1, and No. EP/G056803/1) using HPC resources provided by the TACC at the University of Texas at Austin.


\end{document}